\documentclass[12pt,a4paper]{article}
\pdfoutput=1
\usepackage[latin1]{inputenc}
\usepackage[T1]{fontenc}
\usepackage{amsfonts,amsbsy,bm,euscript,mathrsfs}
\usepackage{amssymb,faktor,slashed}
\usepackage{color}
\usepackage[tbtags]{amsmath}
\usepackage{physics, amsthm, amsmath, tensor, mathtools}
\usepackage[bookmarks=true,colorlinks=true,linkcolor=black,citecolor=black,urlcolor=black,bookmarksnumbered,linktocpage=true]{hyperref}
\usepackage{graphicx}
\usepackage{stringstrings}
\usepackage{xstring}

\newcommand{\rf}[1]{(\ref{#1})}

\usepackage[a4paper,text={170mm,257mm},centering]{geometry}

\numberwithin{equation}{section}

\makeatletter
\renewcommand\section{\@startsection {section}{1}{\z@}%
{-3.5ex \@plus -1ex \@minus -.2ex}%
{2.3ex \@plus.2ex}%
{\normalfont\large\bfseries}}
\renewcommand\subsection{\@startsection{subsection}{2}{\z@}%
{-3.25ex\@plus -1ex \@minus -.2ex}%
{1.5ex \@plus .2ex}%
{\normalfont\normalsize\bfseries}}
\makeatother

\expandafter\def\expandafter\bfseries\expandafter{\bfseries\ifmmode\else\boldmath\fi}
\expandafter\def\expandafter\mdseries\expandafter{\mdseries\ifmmode\else\unboldmath\fi}
\expandafter\def\expandafter\normalfont\expandafter{\normalfont\ifmmode\else\unboldmath\fi}

\providecommand{\href}[2]{#2}

\newcommand{\doilink}[2]{\href{http://doi.org/#2}{#1}}

\newcommand{\mathsym}[1]{{}}
\def\id{\protect{{1 \kern-.28em{\rm l}}}}
\def\be{\begin{eqnarray}}
\def\ee{\end{eqnarray}}

\def\ci{\cite}

\def\a{\alpha}

\def\det{\hbox{det}}

\def\l {\lambda}

\def\const{{\rm const}}

\def\m{\mu}

\def\foot{\footnote}

\def\no{\nonumber}

\def\la{\label}
\def\l{\lambda}

\def\p{\phi}

\def\varpi{{\rm w}}

\def\del{\partial}

\def\ed{\end{document}}

\def\iffa{\iffalse}

\def\vp{\varphi}

\def\ov{\over}

\def\ed{\end{document}}

\def \del{\partial}
\def \a {\alpha}

\def\ov{\over}
\def \ci {\cite}

\def \foot {\footnote}

\def\la{\label}\def\foot{\footnote}

\def \no {\nonumber}

\def \a {\alpha }

\begin{document}


\begin{flushright}\small{Imperial-TP-AT-2020-04  }

\end{flushright}

\vspace{2.5cm}

\begin{center}

{\Large\bf
 Generalised Schwarzschild  metric  from double copy \\
\vspace{0.2cm}
 of  point-like charge solution  in Born-Infeld theory}

\vspace{1.5cm}

{
O. Pasarin\footnote{\ o.pasarin19@ic.ac.uk}  and  A.A. Tseytlin\footnote{\    Also at the Institute for Theoretical and Mathematical Physics of Moscow State University  

\ \ \ \ and Lebedev Institute, Moscow. \  tseytlin@imperial.ac.uk} 
}

\vspace{0.5cm}

{
\em \vspace{0.15cm}
Blackett Laboratory, Imperial College, London SW7 2AZ, U.K.
}
\end{center}

\vspace{0.5cm}

\begin{abstract}
We  discuss possible  application of   the classical double copy  procedure  to   construction of a 
 generalisation of  the Schwarzschild    metric  starting  from   an $\a'$-corrected  open string analogue of the   Coulomb solution.
The latter is   approximated by    a point-like charge   
solution of the Born-Infeld action, which  represents   the  open string effective action  for an abelian vector field 
in the limit  when   derivatives  of the  field strength   are  small. 
  The Born-Infeld  solution has  a regular  electric field   which is  
   constant near the origin   suggesting that  corrections  from  the derivative terms in the open string effective action 
   may be small there.  
    The  generalization of  the Schwarschild  metric obtained by the double copy  construction from 
  the Born-Infeld solution  looks  non-singular  but        the  corresponding    curvature invariants  still   blow up 
     at  $r = 0$.     We  discuss  the origin of this singularity  and comment on possible generalizations. 
\end{abstract}

\newpage

\tableofcontents

\setcounter{footnote}{0}
\setcounter{section}{0}

\section{Introduction}

\ \ \ The (classical) double copy  
is a procedure  to construct gravity solutions from gauge theory ones. 
It originated from the KLT relations  in string theory    and BCJ duality 
 associated to scattering amplitudes in field theory (for a review  see \cite{Bern:2019prr}). 

A  
simple   example  
 of the  double copy is a relation between  the Schwarzschild metric 
  and  the Coulomb potential   $A_{\mu} = (\phi(r), 0, 0, 0)$, $\phi = Q/r $,   created by a point  charge 
       \cite{Monteiro:2014cda}. 
   After a  gauge transformation   
   we get  $A_{\mu} = \phi(r)\,  k_{\mu}$   where 
      $k_{\mu} = (1, ~x_{i}/r)$ is null. 
   Then  the  metric  in the Kerr-Schild form    
   $g_{\mu\nu} = \eta_{\mu\nu}+\phi\, k_{\mu}k_{\nu}$     becomes the 
    Schwarzschild metric  with  mass  $M =2Q$. 

So far almost all  examples  of the double copy    started with  linear Maxwell fields. 
 The  validity and physical  origins of the   classical double   copy   construction at the full non-linear, quantum     
  and string theory   levels are not   
clearly understood at present  but one might speculate that it  may extend beyond the leading order 
in $\a'$ and   relate    exact   open-string and closed-string backgrounds. 
 Here   we will make a first   naive   attempt to study  such an extension.

Gauge theory equations of motion appear  as the leading order  approximation to 
 the effective  field  equations for the massless     vector field  in the  open string theory \ci{Scherk:1974ca}. 
The tree-level open string effective action is given, in the abelian case,  by the Born-Infeld  \cite{Born:1933pep}    term
$\sqrt{ \det ( \eta_{\mu \nu}  + 2 \pi \a'    F_{\mu\nu})}$  \cite{Fradkin:1985qd}
  plus terms depending on derivatives of the field strength $F_{\mu\nu}$ 
 (for a review see  \cite{Tseytlin:1999dj}). 
 We  may   attempt to  first   ignore all  derivative corrections 
 and generalize  the Maxwell's  theory Coulomb solution to its Born-Infeld counterpart \cite{Born:1933pep}.   
    The corresponding electric field is  $E =  Q/\sqrt{r_{0}^{4}+r^{4}}$, 
    where  $r_{0}^{2}= T^{-1}Q$   and   $T= { 1 \ov  2 \pi \alpha'} $ is string tension.  In contrast to the 
     Coulomb  case   here the  field  is non-singular at the origin. This  may be  interpreted  as  
     a consequence  of the inclusion of the   $\alpha'$ corrections  that are expected to "regularize" point-like singularities 
  in string theory   \cite{Tseytlin:1995uq}.
      Since the     field  of the Born-Infeld  solution  is approximately constant near the origin,  this 
      suggests   that  it may be  possible to   consider it  
      as an  approximation to a solution of the  full  (tree-level) 
      open string  effective field equations  in the region close to $r=0$. 
      
One   may wonder whether this regular Born-Infeld  solution  may  double-copy to a generalization of the Schwarzschild metric that  will 
also  be   non-singular at the origin.\foot{One may  argue that to 
 discuss a  possibility of  a double copy for Born-Infeld  fields  one should be 
  assuming  that 
there  exists its non-abelian  version that 
satisfies some form of color/kinematics  duality.} 
Making  the simplest    assumption that  
 the  form   $g_{\mu\nu} = \eta_{\mu\nu}+\phi\, k_{\mu}k_{\nu}$   of the    standard  "leading-order" double copy ansatz  
 is not modified   by the $\a'$-corrections,  the resulting  metric  with the   potential $\phi$ corresponding 
 to the regular  Born-Infeld solution   will look formally non-singular at $r=0$.  However,  as  we will  find  below,  
 the corresponding   curvature  invariants happen to  diverge  at the origin. 
 This   has to do  with too  slow $\phi \sim r$ decay of the scalar  potential at  $r\to 0$. 
  
  We do not  expect   this  singular   $\alpha'$-dependent  double-copy metric   
 to   solve a   closed-string generalization of the Einstein  equations.  
 First,    the  string-theory generalization of the double   copy ansatz  may require  its  non-trivial 
   $\alpha'$-modification.  One  may also need to generalize the  double copy ansatz 
    to  allow  for a  non-zero   dilaton field   \cite{Luna:2016hge,Kim:2019jwm}   which is expected to be non-trivial     for the closed-string 
    generalization of the  Schwarzschild   solution  beyond the leading order in  $\a'$.
    Finally, our use  of the Born-Infeld solution  as an approximation  to the exact open-string solution may be too naive: 
     it is possible   that   (a resummation of)  the derivative corrections in the open-string equations 
    may lead  to a  subtle 
   modification of the Born-Infeld solution  resulting  in  a non-singular  double-copy metric. 

This paper is organised as follows.
 In Section 2  we  will  discuss  the  structure of the open string effective  action 
  and  the  Born-Infeld  solution that we will use.
 In Section 3   we  will  recall   how  the classical double copy procedure may be applied to get the Schwarzschild metric  from 
  the Coulomb   potential.
   In Section 4  we  will  present  the double copy  metric   corresponding to  the   Born-Infeld   solution
    and  discuss  the    singularity of the  corresponding curvature invariants. 
    Section 5  will  contain   some concluding remarks. 
There are  also three technical appendices.

\section{Open string effective action and the Born-Infeld solution} \label{openstringbirewrite}


The effective action  for the abelian gauge  field  in  the  bosonic open string  theory has the following structure 
 \cite{Fradkin:1985qd, Abouelsaood:1986gd, Andreev:1988cb} (we  consider  reduction to  4 dimensions; $T^{-1} \equiv 2\pi\alpha'$)\foot{In superstring case derivative corrections start   with 4-derivative terms.} 
\begin{equation}\la{333}
S
 = c  \int d^{4}x~\sqrt{-\det(\eta_{\mu\nu} +T^{-1}F_{\mu \nu} )}\, \Big[1+T^{-3}\, {B}^{\mu\nu\rho\sigma\lambda\gamma}(T^{-1}F)\, \partial_{\mu}F_{\nu\rho}\partial_{\sigma}\, F_{\lambda\gamma}+\mathcal{O}(\partial^{4}F)\Big],
\end{equation}
where  the  $\del F$-independent part  is the Born-Infeld action   and  
${B}$ is a particular 
function of the field-strength $F_{\mu\nu} = \partial_{\mu}A_{\nu}-\partial_{\nu}A_{\mu}$. 
Explicitly, the  leading order   $\a'^5$ derivative terms are  
 \cite{Andreev:1988cb} 
\begin{align}
S =&c \int d^{4}x  \Big( \sqrt{-\det (\eta+T^{-1}F)}-\frac{1}{48\pi}\hspace{2pt}T^{-5}\Big[ (\partial_{\alpha}F_{\mu\nu})(\partial^{\alpha}F^{\mu\nu})F_{\rho\sigma}F^{\rho\sigma}\no  \\
& + 8(\partial_{\alpha}F_{\mu\nu})(\partial^{\alpha}F^{\nu\lambda})F_{\lambda\rho}F^{\rho\mu} + 4 (\partial_{\alpha}F_{\mu\nu})(\partial^{\beta}F^{\mu\nu})F_{\beta\lambda}F^{\alpha\lambda}\Big] +\mathcal{O}(T^{-7})\Big) \ .\la{666}
\end{align}
 The  resulting   equation for $F_{\mu\nu} $  may be written as:
\begin{align}
& 2\partial_{\mu}\Big[\frac{\partial\sqrt{-\det(\eta+T^{-1}F)}}{\partial F_{\mu\nu}}\Big]-   \frac{1}{12\pi} T^{-5} \Big[(\partial_{\alpha}F_{\lambda\gamma})(\partial^{\alpha}F^{\lambda\gamma})(\partial_{\mu}F^{\mu\nu})\nonumber \\
&\qquad \qquad  +2(\partial_{\mu}\partial_{\alpha}F_{\lambda\gamma})(\partial^{\alpha}F^{\lambda\gamma})F^{\mu\nu}  
 +4\partial_{\mu}\big[(\partial_{\alpha}F_{\sigma\gamma})(\partial^{\mu}F^{\sigma\gamma})F^{\alpha\nu}\big]
 \nonumber \\
 &\qquad \qquad +4 \partial_{\mu}\big[(\partial_{\alpha}F_{\beta\gamma})(\partial^{\alpha}F^{\gamma\mu})F^{\nu\beta}+(\partial_{\alpha}F_{\gamma\lambda})(\partial^{\alpha}F^{\nu\gamma})F^{\lambda\mu}\big] \Big] +\mathcal{O}(T^{-7} )  = 0\ . \label{777}
\end{align}
The Born-Infeld   equation  corresponding to the   vanishing of the    first   term   here    is equivalent to  $ (\eta-T^{-2}F^{2})^{-1}_{~\lambda\mu}~\partial^{\l}F^{\mu\nu} = 0$.
  
  Ignoring  the contributions of the derivative correction terms in (\ref{666})
  let us    look  for a point-like charge     solution of the  Born-Infeld term in (\ref{777}).
In the purely electric case   the  Born-Infeld part of \rf{777}  
reduces to $\partial_{i}\left(E_{i}/\sqrt{1- T^{-2} {E}^{2}}\, \right) =0$. 
If the electric field  is spherically symmetric (corresponding to   a point-like  charge), i.e.  has only 
 the  radial component depending 
  on  $r$  one finds  \cite{Born:1933pep}\foot{For  some applications of this solution see
  \cite{Gibbons:1997xz}.}  
\begin{equation}
E_{r} = F_{0r} = -\partial_{r}A_{0}(r) = \frac{Q}{\sqrt{r_{0}^{4}+r^{4}}}\ , \hspace{10pt}\qquad   \hspace{10pt} r_{0}^{2} \equiv T^{-1}Q. \label{5}
\end{equation}
In contrast to the standard  Coulomb solution  the   Born-Infeld   solution   is regular at $r=0$. 
Since the electric field \rf{5} is   approximately  constant   near near $r=0$,  one 
may hope  that   at least  near the origin    this   background    may  be trusted as a solution to the
 full open string effective action, including  the derivative corrections. 
 A further  discussion of this point    is presented in Appendix \ref{A}. 

Our aim below  will be  to  construct the  double copy metric corresponding to the scalar potential in \rf{5} 
that generalizes the Schwarzschild metric   which is the double copy of the Coulomb potential.


\def \vp {\varphi}

\section{Schwarzschild metric  as the double copy   of  the Coulomb solution }\label{schwarzschild}

\hspace{20pt} Let us  first  briefly  review 
 the  application of the classical double copy procedure to the Schwarzschild   solution \cite{Monteiro:2014cda}. 
 The Schwarzschild metric is a particular case of
\begin{equation}
ds^{2} = -\big[1- \phi (r) \big]dt^{2}+\frac{dr^{2}}{1-\phi(r) }+r^{2}(d\theta^{2}+\sin^{2}\theta \hspace{2pt}d\vp^{2})\ , \label{02}
\end{equation}
with  $\phi = 2M/r$. 
Changing coordinates to $(\bar{t}, x_i)$,\  $\bar{t} \equiv t+2M\ln(r-2M)$,  the Schwarzschild metric
 can be written in the Kerr-Schild  form 
 \begin{equation}\la{01}
g_{\mu\nu} = \eta_{\mu\nu}+\phi\,  k_{\mu}k_{\nu}\ , \qquad \qquad 
 k_{\mu} \equiv \big(1, \frac{x^{i}}{r}\big)\ , \qquad k_\m k^\mu =0 \ . 
\end{equation}
 This   may be interpreted as   a double copy 
 corresponding to an abelian  gauge potential 
 \begin{equation}\la{10}
A_{\mu} = \phi(r) \, k_{\mu} \ ,  \qquad \qquad  \phi = \frac{Q}{ r} \ .
\end{equation}
assuming   that   $ Q\equiv 2M$.\foot{We shall   ignore normalization 
 constants in the definition of mass and charge.}
 The  potential  \rf{10}  is gauge-equivalent to the Coulomb potential $A_\mu = \phi ( 1, 0,0,0)$

 For general $\p(r)$,  the  change of coordinates  bringing    the metric \rf{02} to  the  Kerr-Schild form \rf{01} 
  can be found  by looking for  radial null geodesics  of  \eqref{02}.
 Setting  
$- (1-\phi)dt^{2}+\frac{dr^{2}}{1-\phi} = 0$  gives 
 the following integral  representation for $t$ (denoted by  $t^{*}(r)$):
\begin{equation}
t^{*}(r) = \pm\int\frac{dr}{1-\phi(r)}\ .
\end{equation}
In the Schwarzschild case of  $\phi = 2M/r$  this  gave   $t^{*}(r) = r+2M\ln(r-2M)$.
The Kerr-Schild form of \rf{02} 
 is then obtained by changing  from $(t, r, \theta, \phi)$   to $(\bar{t}, x_i )$  coordinates 
  where $x_i$ are the standard cartesian ones   and  $\bar{t} \equiv t-r+t^{*}(r)$. 
  To   perform   the change of  coordinates 
it is sufficient to  use  the differential  of 
$\bar{t} = t-r+t^{*}$, i.e.  
\begin{equation}
d\bar{t} = dt+\frac{\phi(r)}{1-\phi(r)}\hspace{2pt}dr. \label{222}
\end{equation}

\section{Double copy of  the Born-Infeld solution}\label{doublebi}

\hspace{20pt} To  construct  the classical double copy metric for  the Born-Infeld   solution  in \eqref{5}
 we  need   the corresponding gauge potential $A_{\mu}$. Integrating \eqref{5} over $r$  with the boundary condition 
 $A_0 \big|_{r\to \infty} \to 0$ gives 
\begin{equation}\label{1}
A_{0}(r) \equiv \phi(r) = \int_{r}^{\infty} dr' \, E_r (r') = \frac{Q}{r}\hspace{2pt} \prescript{}{2}{F}_{1}\Big(\frac{1}{4}, ~\frac{1}{2}, ~\frac{5}{4}, -\frac{r_{0}^{4}}{r^{4}}\Big) =  \frac{Q}{r} \Big[ 1 - {r_0^4\ov 10 r^4}   + \mathcal{O}({r_0^{8}\ov r^8})\Big] \ , 
\end{equation}
where $ {}_2F_1$ is the standard hypergeometric function. 
By a gauge transformation    $A_\mu = (\phi , 0, 0, 0)$  can be transformed   into (cf. \rf{10})  
\begin{equation}\label{7}
A_{\mu} = \phi(r)\, k_{\mu} = \phi(r)\left(1, ~\frac{x_{i}}{r}\right) \ .  
\end{equation}   The corresponding 
 double copy metric is  then \rf{01}   with   $\phi(r)$    given by   \rf{1}. 
Here   $ds^{2} = g_{\mu\nu}(x) dx^{\mu}dx^{\nu}$ with $x^{\mu} = (\bar{t}, x_i)$  and 
 $\bar{t}$  related to $t$ as  in \rf{222}. 
 Using this relation and the  transformation  between the cartesian and the spherical coordinates
 we find  that the metric  takes the same  "Schwarschild"  form \rf{02}  now with  Coulomb $\p= { Q \ov r}$ replaced by 
  $\p(r)$ in \rf{1}. It   thus generalizes the Schwarschild   metric to the case  when $r_0^2 = 2 \pi \alpha' Q$
  is non-zero. 
 
In contrast to the Schwarschild  metric the components of the  resulting metric \eqref{02}   look 
non-singular    since 
$\phi(r)$  in (\ref{1})     has a regular expansion for small $r$:
\begin{equation}
\phi(r) = c_0 + c_1 r + c_5 r^5 + \mathcal{O}(r^{9}) = 
 { Q \ov r_0} \Gamma(\tfrac{5}{4}) \Big[ { \Gamma(\tfrac{1}{4})}{\sqrt{\pi}}-\frac{r}{\Gamma(\tfrac{1}{4})\, r_0}
+\frac{r^5 }{5r_{0}^{5}}\Big] +\mathcal{O}(r^{9})\ .  \label{2}
\end{equation}
Somewhat surprisingly,  the corresponding  curvature invariants  still   turn out  to be  singular at $r=0$. 
For example,  the scalar curvature  is   given by 
\begin{equation}\la{22}
R = \frac{2\phi(r)}{r^{2}} -  \frac{2Q(r^{4}+2r_{0}^{4})}{r (r^{4}+r_{0}^{4})^{3/2} } 
= \frac{2Q\hspace{2pt}\Gamma(\tfrac{1}{4})\Gamma(\tfrac{5}{4})}{r_{0}\sqrt{\pi}}\hspace{2pt}\frac{1}{r^{2}}-\frac{4Q[\Gamma(\tfrac{1}{4})+2\Gamma(\tfrac{5}{4})]}{r_{0}^{2}\hspace{2pt}\Gamma(\tfrac{1}{4})}\hspace{2pt}\frac{1}{r}+\mathcal{O}(r^{3}) \ . 
\end{equation}
This  singularity is due to the presence  of  the  first two ($c_0$ and $c_1 r$) terms in the $r\to 0$   expansion of $\p$ in \rf{2}.

If $\phi\big|_{r\to 0} = c_0\not=0$  then the metric \rf{02}  has a conical singularity at $r=0$. 
This is  not, however,   a serious issue as 
  we can set $c_0=0$  by  changing the integration   constant in \rf{1}  (or by a gauge transformation of the potential \rf{7})
  and then define the double copy  metric  \rf{01}   using  $\phi$ in this   gauge.\foot{In this case, however, 
   instead of $\phi\big|_{r\to \infty} =0$ we will have   $\phi\big|_{r\to \infty} =- c_0$     
   so that will change the standard  Minkowski asymptotic form of the  metric \rf{02}.}
The real problem is  that  $\phi\big|_{r\to 0} = c_1 r= c_1 \sqrt{ x_i^2}  $    is  non-analytic in cartesian coordinates 
and this  effectively    produces   singularity  in the curvature invariants. 
This $c_1 r$ term   can not be eliminated   by a  gauge transformation as it  is   responsible 
for   the  non-zero constant    value of the Born-Infeld 
  electric field  $E_{r} =- \partial_{r} \phi$ in  \rf{5}  at $r=0$   (see also  Appendix C).

In general, if  we start with the    metric (\ref{02})  with  $\phi$ 
having a   regular  $r\to 0$  expansion  
\begin{equation}\label{3} 
\phi(r) = c_{0}+c_{1}r+c_{2}r^{2}+c_{3}r^{3}+c_{4}r^{4}+c_{5}r^{5}+\mathcal{O}(r^{6}), 
\end{equation}
then the $r\to 0$ expansion   of    the  curvature squared   invariant is  found to be 
\begin{align}
R_{\mu\nu\rho\sigma}R^{\mu\nu\rho\sigma}\big|_{r\to 0}  = & \hspace{4pt}\frac{4c_{0}^{2}}{r^{4}}+\frac{8c_{0}c_{1}}{r^{3}}+\frac{4c_{1}^{2}+8c_{0}c_{2}}{r^{2}}+\frac{8(c_{1}c_{2}+c_{0}c_{3})}{r} \\ \qquad 
& +4(c_{2}^{2}+2c_{1}c_{3}+2c_{0}c_{4})+8(c_{2}c_{4}+c_{1}c_{4}+c_{0}c_{5})r+\mathcal{O}(r^{2})\ .  \label{eq:r2}
\end{align}
Thus  it is   non-vanishing   $c_0$ and $c_1$  that are,  indeed, responsible  for the  singularity. 
Explicitly, in the case of $\p(r)$ in \rf{2} we find (see also  \rf{77}) 
\begin{align}
&R_{\mu\nu\rho\sigma}R^{\mu\nu\rho\sigma}   =\frac{4[\phi(r)]^{2}}{r^{4}}   + \frac{8Q^{2}(r^{8}+r_{0}^{4}r^{4}+\tfrac{1}{2} r_{0}^{8})}{r^{2}(r^{4}+r_{0}^{4})^{3}}   = 
\frac{4Q^{2}\hspace{2pt}\Gamma(\tfrac{1}{4})^{2}\Gamma(\tfrac{5}{4})^{2}}{\pi r_{0}^{2}}\hspace{2pt}\frac{1}{r^{4}}
\no  \\
&\qquad \qquad \qquad  -\frac{32Q^{2}\Gamma(\tfrac{5}{4})^{2}}{\sqrt{\pi}\hspace{2pt}r_{0}^{3}}\hspace{2pt}\frac{1}{r^{3}}+\frac{4Q^{2}}{r_{0}^{4}}\frac{\Gamma(\tfrac{1}{4})^{2}+16\Gamma(\tfrac{5}{4})^{2}}{\Gamma(\tfrac{1}{4})^{2}}\hspace{2pt}\frac{1}{r^{2}}  +\frac{16Q^{2}\hspace{2pt}\Gamma(\tfrac{5}{4})^{2}}{5r_{0}^{7}\sqrt{\pi}}\hspace{2pt}r
+\mathcal{O}(r^{3})\ . \
\la{88}
\end{align}
 Note   that    the expressions  in \rf{22}  and \rf{88} (before expanding near  $r\to 0$) reduce to the standard
  Schwarzschild 
  values ($ R=0, \ R_{\mu\nu\rho\sigma}R^{\mu\nu\rho\sigma}  =  { 12 Q^2 \ov r^6}$) 
  once we set $r_0=  2 \pi \a' Q =0$  for fixed $r$.  For  non-zero  $r_0$ the 
  corresponding metric  \rf{02} has  a non-trivial Ricci tensor
  (see   Appendix \ref{B}).
  It is not  clear if there is  some generalization of the Einstein 
  equations  for which the  metric \rf{02} with $\p$ in \rf{1}  is a solution.

\section{Concluding remarks}

\hspace{20pt} 
Our aim  in this note  was to explore if   the  simplest  classical double copy  ansatz   may produce a non-singular generalization 
of the Schwarzschild metric if applied to the exact    open-string   analog  of the Coulomb solution. The latter was 
  assumed to be  approximated   by the Born-Infeld   solution.  
We   suggested   that since the Born-Infeld   action 
is the leading term in the open string effective action  expansion in powers of  field strength derivatives 
 and since  the  electric field   of the Born-Infeld  analog of the Coulomb solution 
is  approximately   constant   near $r=0$   this  solution may be  trusted  near the origin. 

The resulting double copy   metric reduces to the Schwarzschild one in the 
$\a'\to 0$ limit  
 and at first sight  seems regular     near $r=0$. However, the decay of the  Born-Infeld scalar 
 potential $\phi$   for     $r\to 0$  
 happens to be  too slow 
(reflecting  the  non-vanishing    value of the Born-Infeld field at the origin) 
for the  corresponding  curvature  invariants  to be regular.  This  does not 
 of course imply  the  singularity of a 
closed string generalization of the Schwarzschild   solution since 
there is no a priori reason to expect   this double copy  construction to produce 
  a solution  of the   closed string effective   equations   and also   given   that  the Born-Infeld  field \rf{5}
   is not an exact solution of the open-string theory.

One direction   to  investigate  further  is the influence  of derivative   corrections  in the open string  effective action
 on the behaviour 
of the  corresponding solution near $r=0$,   going beyond a  simplified  analysis in Appendix A.  In  particular,  one 
 may wonder if a resummation of derivative corrections  may alter  the $\phi\big|_{r\to 0}  \to r$   behaviour   of the  scalar potential
 that   may 
 resolve the singularity  of the double copy metric. 
It is also  interesting to study  a possible generalization  
of the  double copy ansatz \cite{Kim:2019jwm}
that allows for a non-trivial dilaton. 
More generally,  the status  of the double copy construction  beyond the   leading order in $\alpha'$ expansion 
 and whether  it  
 may provide a 
 map  between  exact    open-string and closed-string solutions  
   remains to be explored.

\section*{Acknowledgments}
We are grateful to Tim Adamo and Radu Roiban    for very useful comments on the draft.  
This work was supported  by  the 
 STFC grant ST/P000762/1.

\bigskip
\newpage
\appendix

\section{Born-Infeld solution as an approximation to open-string solution} \label{A}
\def\theequation{A.\arabic{equation}}
\setcounter{equation}{0}

\ \ \ Assuming the same ansatz  for $F_{\mu \nu}$ (no magnetic field, time-independent electric field) 
  that  led to the  Born-Infeld solution  in \eqref{5}, only 
  the $\nu = 0$ component of the equations  \eqref{777} is non-trivial  and  may be written as (ignoring  higher order     terms in \rf{777})
\begin{align}
&\del_i \Big(\frac{{E_i}}{\sqrt{1- T^{-2} {E}^{\hspace{1pt}2}}}\Big)  + \frac{1}{6\pi T^{3}}\Big[  \frac{1}{2}\hspace{2pt}(\partial_{k}E_{i})(\partial_{k}E_{i})\del_j E_j +(\del_i\del_j E_k)(\del_j E_k) E_i\nonumber \\[10pt] & +2\del_j [(\del_i E_k)(\del_j E_k)E_i]
 +\del_i [(\del_j E_k)(\del_j E_i)E_k]+\del_i [(\del_j E_k)(\del_j E_k)E_i]\Big] = 0. \label{eq:eom0re}
\end{align}
Assuming further  that $E_i$ is spherically-symmetric   we   get  ($E\equiv E_r(r) $)
\begin{equation}
\partial_{r}\Big[\frac{r^{2}E}{\sqrt{1-T^{-2}E^{2}}}\Big]+ 
 \frac{3}{4\pi T^3}  \partial_{r}E  \Big[2r E^{2}+r^{3}(\partial_{r}E)^{2}+2r^{2}E(\partial_{r}E+r\partial_{r}^{2}E)\Big] = 0\ . \label{eq:A2}
\end{equation}
From here we may   find the leading  correction to the Born-Infeld   solution   coming from the presence of the field strength derivative terms 
in the open string effective action. Setting    $E(r) = E^{(0)}(r)+E^{(1)}(r)$, where $E^{(0)}(r)$ is the Born-Infeld solution \rf{5}
  we obtain  from  \eqref{eq:A2}  the following first-order differential equation for $E^{(1)}$\ \  ($r^2_0=T^{-1} Q$):
\begin{equation}
\frac{dE^{(1)}}{dr}+\frac{2(r^{4}-2r_{0}^{4}) }{r (r^4 + r_0^4) }\, E^{(1)}= \frac{3r_0^6 r^7(7r^{8}-6r_{0}^{4}r^{4}+r_{0}^{8})}{\pi (r^{4}+r_{0}^{4})^{6}} \ . 
\end{equation}
Its  solution may be written as:
\begin{equation}
E^{(1)} = -\frac{ r_0^6 r^{4}(7r^{8}+2r_{0}^{4}r^{4}+r_{0}^{8})}{2\pi (r^{4}+r_{0}^{4})^{5}} = -\frac{r^4}{2\pi r_{0}^{6}}\hspace{2pt}+\frac{3r^8}{2\pi r_{0}^{10}}\hspace{2pt}+\mathcal{O}(r^{12}) \ .
\end{equation}
Its expansion  for  $r\to 0$  starts  at order $r^{4}$  so  it does not   change    the $E\big|_{r\to 0}= {Q\ov r^2_0} = \const$ 
behaviour of the   Born-Infeld   field \rf{5}  near the origin,  suggesting it can be trusted  near  $r=0$. 
Equivalently, the derivative terms  do not alter the leading $c_1 r $ term in the scalar potential \rf{2}  that was found to be   responsible for the singularity of the double-copy metric.

\iffa 
, so the contribution of this correction near $r = 0$ is small and the Born-Infeld solution may hold near the origin even when leading-order derivative terms are included. Whether this statement holds at all orders in field-strength derivatives is not clear. Since the expansion of the Born-Infeld solution is $E^{(0)} = a+br^{4}+\ldots$, third-order derivatives in the action, or equivalently, fourth-order derivatives in the equations of motion may give a significant contribution. In the case of the superstring (eq. 7.3 in \cite{Tseytlin:1999dj}) such four-derivative terms are absent from the equations of motion, so the Born-Infeld solution would still be valid past those corrections.  
\fi

\section{Curvature  tensor for the double copy metric} \label{B}
\def\theequation{B.\arabic{equation}}
\setcounter{equation}{0}

\ \ \ The  curvature tensor  for the  metric of the form \eqref{02}  can be  computed 
for general  function  $\phi (r)$  with the  non-trivial  components  being \begin{align}
R\indices{^{t}_{rtr}} = & \hspace{4pt}\frac{\phi''}{2(1-\phi)} , \qquad 
R\indices{^{t}_{\theta\theta t}} =  \hspace{4pt}R\indices{^{r}_{\theta\theta r}} 
= -\frac{r}{2}\hspace{2pt}\phi' ,\qquad 
R\indices{^{t}_{\vp\vp t}} =  \hspace{4pt}R\indices{^{r}_{\vp\vp r}} = R\indices{^{t}_{\theta\theta t}}\sin^{2}\theta,  \nonumber \\
R\indices{^{r}_{ttr}} = & \hspace{4pt}\frac{1}{2}\hspace{2pt}(1-\phi)\hspace{2pt}\phi'' ,\qquad 
R\indices{^{\theta}_{tt\theta}} =  \hspace{4pt}R\indices{^{\vp}_{tt\vp}} = \frac{1-\phi}{2r}\hspace{2pt}\phi'
 , \qquad R\indices{^{\theta}_{r\theta r}} =  \hspace{4pt}R\indices{^{\vp}_{r\vp r}} = \frac{\phi'}{2r(1-\phi)} 
, \nonumber  \\[10pt]
R\indices{^{\theta}_{\vp\theta\vp}} = & \ \ R\indices{^{\vp}_{\theta\vp\theta}}\sin^{2}\theta = \phi(r) \sin^{2}\theta.
\end{align}
For the Ricci tensor  and scalar  we get:
\begin{align}
R_{tt} = - \hspace{4pt}\frac{1-\phi}{2r}\hspace{2pt}(2\phi'+r\phi''), \qquad  &
R_{rr} =  \hspace{4pt}\frac{2\phi'+r\phi''}{2r-2r\phi}, \qquad R_{\theta\theta}  =
{R_{\vp\vp}\over \sin^{2}\theta}= \phi+r\phi' \ , \\
&
R = \frac{2\phi}{r^{2}}+\frac{4\phi'}{r}+\phi'' \ . \label{eq:r2general}
\end{align}
The explicit form of the 
  Ricci tensor corresponding to the metric  \eqref{02}  with $\phi$ in \eqref{1} is 
\begin{align}
R_{tt} = & \hspace{4pt}\frac{Q\hspace{2pt}r_{0}^{4}\hspace{2pt}(1-\phi)}{r^7(1+{r_{0}^{4}\ov r^4})^{3/2}}
 = \frac{Q\hspace{2pt}r_{0}\hspace{2pt}\sqrt{\pi}-Q^{2}\hspace{2pt}\Gamma(\tfrac{1}{4})\Gamma(\tfrac{5}{4})}{r_{0}^{3}\hspace{2pt}\sqrt{\pi}}\hspace{2pt}\frac{1}{r}+\frac{4Q^{2}\hspace{2pt}\Gamma(\tfrac{5}{4})}{r_{0}^{4}\Gamma(\tfrac{1}{4})}+\mathcal{O}(r^{3}), \no \\[10pt]
R_{rr} = & \hspace{4pt}\frac{Q\hspace{2pt}r_{0}^{4}}{(1-\phi)(1+ {r_{0}^{4}\ov r^4})^{3/2}}=-\frac{Q\hspace{2pt}\sqrt{\pi}}{r_{0}[r_{0}\sqrt{\pi}-Q\hspace{1pt}\Gamma(\tfrac{1}{4})\Gamma(\tfrac{5}{4})]}\hspace{2pt}\frac{1}{r}+\frac{4Q^{2}\pi\hspace{2pt}\Gamma(\tfrac{5}{4})\Gamma(\tfrac{1}{4})}{r_{0}^{2}[r_{0}\sqrt{\pi}-Q\hspace{2pt}\Gamma(\tfrac{1}{4})\Gamma(\tfrac{5}{4})]^{2}}+\mathcal{O}(r) ,\no  \\[10pt]
R_{\theta\theta} = & { R_{\vp\vp}  \over  \sin^{2}\theta}  = \hspace{4pt}\phi(r)-\frac{Q}{r}\Big(1+\frac{r_{0}^{4}}{r^{4}}\Big)^{-1/2} = \frac{Q\hspace{2pt}\Gamma(\tfrac{1}{4})\Gamma(\tfrac{5}{4})}{r_{0}\sqrt{\pi}}-\frac{r_{0}^{2}\hspace{2pt}\Gamma(\tfrac{1}{4})+8Q\Gamma(\tfrac{5}{4})}{2r_{0}^{2}\hspace{2pt}\Gamma(\tfrac{1}{4})}\hspace{2pt}r+\mathcal{O}(r^{3}). \end{align}
The curvature squared   invariant is 
\begin{equation}
R_{\mu\nu\rho\sigma}R^{\mu\nu\rho\sigma} = \frac{4\phi^{2}}{r^{4}}  +\frac{4\phi'^2}{r^2}+\phi''^{2}
=\frac{4\phi^{2}}{r^{4}}   +  \frac{8Q^{2}(r^{8}+r_{0}^{4}r^{4}+\tfrac{1}{2} r_{0}^{8})}{r^{2}(r^{4}+r_{0}^{4})^{3}}\ ,  \label{77}
\end{equation}
with its  expansion at $r\to 0$ given in \rf{88}. 
The  Weyl tensor  squared   is   also singular at $r\to 0$
\begin{align}
C_{\mu\nu\rho\sigma}C^{\mu\nu\rho\sigma} =  \frac{(2\phi-2r\phi'+r^{2}\phi'')^{2}}{3r^{4}}  
= \frac{4Q^{2}\Gamma(\tfrac{1}{4})^{2}\Gamma(\tfrac{5}{4})^{2}}{3\pi r_{0}^{2}}\hspace{2pt}\frac{1}{r^{4}}+\mathcal{O}(r^{-3}).
\end{align}

\section{Gauge transformation of the vector potential  near  $r=0$  } \label{C}
\def\theequation{C.\arabic{equation}}
\setcounter{equation}{0}

\ \ \ Given  the vector  potential 
  $A_{\mu} = \phi(r)(1, ~x_{i}/r)$ with $\phi \big|_{r\to 0 } = c_{0}+c_{1}r+c_{5}r^{5} +... $   as in \eqref{7},\eqref{2}, 
let us  see if  there is  a gauge transformation  that eliminates $c_0$ and $c_1$ terms, i.e. if $A_{\mu} $
can be transformed  into  
\begin{equation}
\tilde{A}_{\mu}  
 = \tilde{\phi}(r)\big(1, \frac{x_{i}}{r}\big),\qquad \qquad  \tilde \phi(r) \big|_{r\to 0 } = \tilde c_{5}r^{5} +...\ . 
\end{equation}
The relation $\tilde{A}_{\mu} =  A_{\mu}-\partial_{\mu}\chi$ implies 
\begin{equation}
\partial_{0}\chi = c_{0}+c_{1}r+(c_{5}-\tilde{c}_{5})r^{5} + ... ,\qquad   \hspace{10pt} \partial_{i}\chi = \frac{c_{0}x_{i}}{r}+c_{1}x_{i}+(c_{5}-\tilde{c}_{5})x_{i}r^{4} + ...\ . \label{eq:gtransf}
\end{equation}
These equations   lead to   
\begin{align}
\chi(t, x) = & \big[c_{0}+c_{1}r+(c_{5}-\tilde{c}_{5})r^{5}\big]\, t+f(x)\  , \\
\partial_{i}f(x) = &\frac{c_{0}x_{i}}{r}+c_{1}x_{i}+(c_{5}-\tilde{c}_{5})x_{i}r^{4}-\left[\frac{c_{1}x_{i}}{r}+5(c_{5}-\tilde{c}_{5})x_{i}r^{3}\right]t.\label{11} 
\end{align}
The left-hand side of  \eqref{11}  is time-independent,  
so  it is consistent only if     $c_{5} = \tilde{c}_{5}$ and $c_{1} = 0$. 
Thus $c_1$ cannot be eliminated  by a  gauge transformation.

\newpage


\end{document}

\bibitem{Kawai:1985xq} 
H.~Kawai, D.~C.~Lewellen and S.~H.~H.~Tye,
``A Relation Between Tree Amplitudes of Closed and Open Strings,''
Nucl.\ Phys.\ B {\bf 269}, 1 (1986).

\bibitem{Bern:2008qj} 
Z.~Bern, J.~J.~M.~Carrasco and H.~Johansson,
``New Relations for Gauge-Theory Amplitudes,''
Phys.\ Rev.\ D {\bf 78}, 085011 (2008)
[arXiv:0805.3993 [hep-ph]].

\bibitem{Bern:2010yg} 
Z.~Bern, T.~Dennen, Y.~t.~Huang and M.~Kiermaier,
``Gravity as the Square of Gauge Theory,''
Phys.\ Rev.\ D {\bf 82}, 065003 (2010)
[arXiv:1004.0693 [hep-th]].

\bibitem{[KerrSchild]} R.~P. ~Kerr, A. ~Schild, ``Republication of: A new class of vacuum solutions of the Einstein field equations,'' \doilink{Gen Relativ Gravit 41, 2485?2499 (2009)}{https://doi.org/10.1007/s10714-009-0857-z}.

\bibitem{Monteiro:2014cda} 
R.~Monteiro, D.~O'Connell and C.~D.~White,
``Black holes and the double copy,''
JHEP {\bf 1412}, 056 (2014)
[arXiv:1410.0239 [hep-th]].

\bibitem{Bah:2019sda} 
I.~Bah, R.~Dempsey and P.~Weck,
``Kerr-Schild Double Copy and Complex Worldlines,''
JHEP {\bf 2002}, 180 (2020)
[JHEP {\bf 2020}, 180 (2020)]
[arXiv:1910.04197 [hep-th]].

\bibitem{LunaGodoy:2018tyq} 
A.~Luna Godoy,
``The double copy and classical solutions.''
\bibitem{Kim:2019jwm} 
K.~Kim, K.~Lee, R.~Monteiro, I.~Nicholson and D.~Peinador Veiga,
``The Classical Double Copy of a Point Charge,''
JHEP {\bf 2002}, 046 (2020)
[arXiv:1912.02177 [hep-th]].

\bibitem{Fradkin:1985qd} 
E.~S.~Fradkin and A.~A.~Tseytlin,
``Nonlinear Electrodynamics from Quantized Strings,''
Phys.\ Lett.\  {\bf 163B}, 123 (1985).

\bibitem{Andreev:1988cb} 
O.~D.~Andreev and A.~A.~Tseytlin,
``Partition Function Representation for the Open Superstring Effective Action: Cancellation of Mobius Infinities and Derivative Corrections to Born-Infeld Lagrangian,''
Nucl.\ Phys.\ B {\bf 311}, 205 (1988).

\bibitem{Gibbons:1997xz} 
G.~W.~Gibbons,
``Born-Infeld particles and Dirichlet p-branes,''
Nucl.\ Phys.\ B {\bf 514}, 603 (1998)
[hep-th/9709027].

\bibitem{Abouelsaood:1986gd} 
A.~Abouelsaood, C.~G.~Callan, Jr., C.~R.~Nappi and S.~A.~Yost,
``Open Strings in Background Gauge Fields,''
Nucl.\ Phys.\ B {\bf 280}, 599 (1987).

\end{thebibliography}

\end{document}

\bibitem{Tseytlin:1987ww} 
A.~A.~Tseytlin,
``Renormalization of Mobius Infinities and Partition Function Representation for String Theory Effective Action,''
Phys.\ Lett.\ B {\bf 202}, 81 (1988).
\bibitem{Gonorazky:1998vk} 
S.~Gonorazky, C.~Nunez, F.~A.~Schaposnik and G.~A.~Silva,
``Bogomol'nyi bounds and the supersymmetric Born-Infeld theory,''
Nucl.\ Phys.\ B {\bf 531}, 168 (1998)
[hep-th/9805054].

\bibitem{Thorlacius:1997zd} 
L.~Thorlacius,
``Born-Infeld string as a boundary conformal field theory,''
Phys.\ Rev.\ Lett.\  {\bf 80}, 1588 (1998)
[hep-th/9710181].

\bibitem{Brecher:1998tv} 
D.~Brecher,
``BPS states of the non Abelian Born-Infeld action,''
Phys.\ Lett.\ B {\bf 442}, 117 (1998)
[hep-th/9804180].

\bibitem{Suzuki:1997we} 
T.~Suzuki,
``Born-Infeld action in (n + 2)-dimension, the field equation and a soliton solution,''
Lett.\ Math.\ Phys.\  {\bf 47}, 159 (1999)
[hep-th/9712002].

\bibitem{Yang:2000uj} 
Y.~S.~Yang,
``Classical solutions in the Born-Infeld theory,''
Proc.\ Roy.\ Soc.\ Lond.\ A {\bf 456}, 615 (2000).

\bibitem{Kruglov:2016uzf} 
S.~I.~Kruglov,
``Notes on Born-Infeld-type electrodynamics,''
Mod.\ Phys.\ Lett.\ A {\bf 32}, no. 36, 1750201 (2017)
[arXiv:1612.04195 [physics.gen-ph]].

\bibitem{Tseytlin:1996it} 
A.~A.~Tseytlin,
``Self-duality of Born-Infeld action and Dirichlet three-brane of type IIB superstring theory,''
Nucl.\ Phys.\ B {\bf 469}, 51 (1996)
[hep-th/9602064].

\bibitem{Perry:1996mk} 
M.~Perry and J.~H.~Schwarz,
``Interacting chiral gauge fields in six-dimensions and Born-Infeld theory,''
Nucl.\ Phys.\ B {\bf 489}, 47 (1997)
[hep-th/9611065].